\documentclass{aip-cp}

\usepackage[numbers]{natbib}
\usepackage{rotating}
\usepackage{graphicx}

\usepackage{xspace}

\newcommand{\htp}{H$_2^+$\xspace}
\newcommand{\nuebar}{$\bar{\nu}_e$\xspace}

\begin{document}

\title{First Commissioning Results of the Multicusp Ion Source at MIT (MIST-1) for \htp}

\author[aff1]{D. Winklehner\corref{cor1}}
\author[aff1]{S. Axani}
\author[aff2]{P. Bedard}
\author[aff1]{J. Conrad}
\author[aff1]{J. Corona}
\author[aff1]{F. Hartwell}
\author[aff1]{J. Smolsky}
\author[aff1]{A. Tripathee}
\author[aff1]{L. Waites}
\author[aff3]{P. Weigel}
\author[aff1]{T. Wester}
\author[aff4]{M. Yampolskaya}

\affil[aff1]{Massachusetts Institute of Technology, Cambridge, MA 02139, USA}
\affil[aff2]{Reed College, Portland, OR 97202, USA}
\affil[aff3]{Drexel University, Philadelphia, PA 19104, USA}
\affil[aff4]{Cornell University, Ithaca, New York 14853, USA}

\corresp[cor1]{Corresponding author: winklehn@mit.edu}

\date{\today}
%$\ce{^{9}\mathrm{Be}}$ 
%$\ce{^{7}\mathrm{Li}}$
\maketitle
\begin{abstract}
IsoDAR is an experiment under development to search for sterile neutrinos using the isotope Decay-At-Rest (DAR) production mechanism, where protons impinging on $^9$Be create neutrons which capture on 
$^7$Li which then beta-decays producing \nuebar. 
As this will be an isotropic source of \nuebar, the primary driver current must be large (10 mA cw) for IsoDAR to have sufficient statistics to be conclusive within 5 years of running. \htp was chosen as primary ion to overcome some of the space-charge limitations during low energy beam transport and injection into a compact cyclotron. The \htp will be stripped into protons before the target. At MIT, a multicusp ion source (MIST-1) was designed and built to produce a high intensity beam with a high \htp fraction. MIST-1 is now operational at the Plasma Science and Fusion Center (PSFC) at MIT and under commissioning. 
\end{abstract}

\section{INTRODUCTION}
\begin{figure}[!b]
	\centering
		\includegraphics[width=0.65\textwidth]
        {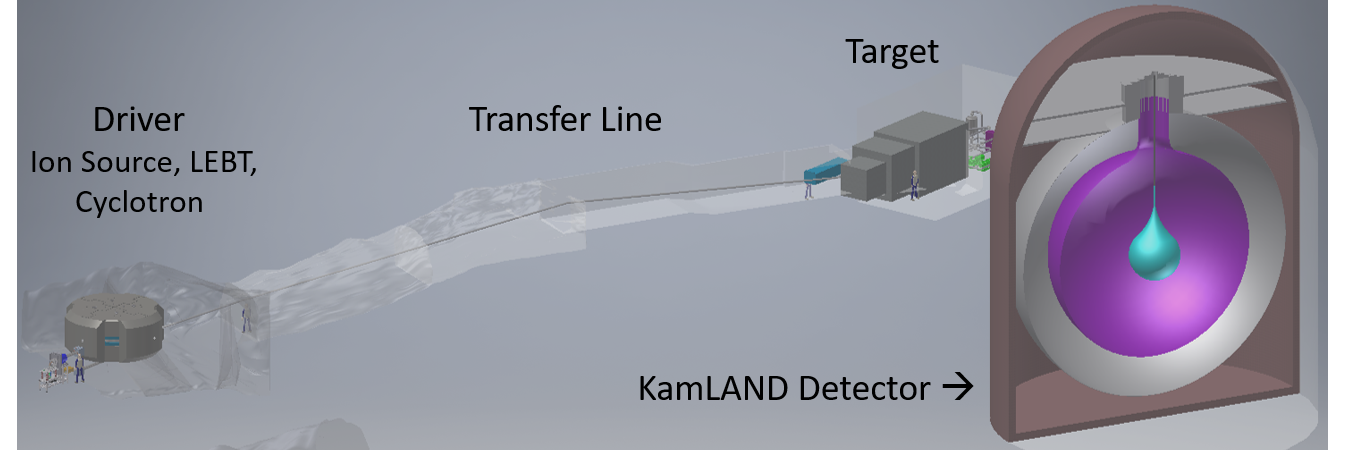}
	    \caption{Cartoon of the IsoDAR experiment paired with the KamLAND detector. 
	             \htp is produced in 
	             the MIST-1 ion source, injected into a cyclotron through
	             a spiral inflector, accelerated to 
	             60 MeV/amu , extracted, stripped into protons and 
	             transported to the neutrino production target. 
	             CAD model courtesy of Larry Bartoszek. (Color online.)}
	\label{fig:isodar}
\end{figure}
The IsoDAR (Isotope Decay-At-Rest) experiment is designed to measure the disappearance of \nuebar in the 10 MeV range over a short baseline of 16 m
\cite{adelmann:isodar}. 
Comparing the predicted survival rate with the measured one, IsoDAR will 
be able to test the sterile neutrino hypothesis and distinguish between 
models with one and two extra neutrinos that are not in the standard model.
Neutrinos are produced isotropically on a target through beta-decay-at-rest 
of $^8$Li, which in turn is created by neutron capture on very pure (99.99\%) $^7$Li. To achieve the necessary \nuebar flux in the detector, a very high 
primary proton current (10 mA cw) is required which will be delivered by
a compact 60 MeV/amu isochronous cyclotron. To overcome space charge issues
during injection and capture, \htp is being accelerated and stripped into
protons after extraction from the cyclotron. A cartoon of the IsoDAR experiment
set up in the Kamioka mine in Japan is depicted in Fig. \ref{fig:isodar}.

Experimental studies of the production of a high intensity \htp ion beam
and injection into a cyclotron were performed \cite{winklehner:bcs_tests}
using the off-resonance flat-field ECR ion source VIS (versatile ion
source) \cite{miracoli:vis1}. The results suggested that the final goal of
extracting 5 mA of \htp from the cyclotron was possible, albeit with 
difficulty, due to a slightly too low initial \htp beam current. 
Two upgrade routes were conceived: MIST-1, a dedicated 
ion source for \htp \cite{axani:mist1}, and better pre-bunching with a Radio Frequency Quadrupole (RFQ), directly injecting beam into the cyclotron
\cite{winklehner:rfq2}. Both are being pursued. In this paper we report on the status of the MIST-1 ion source.

\section{ION SOURCE DESIGN}
\begin{figure}
	\centering
		\includegraphics[width=0.4\textwidth]
        {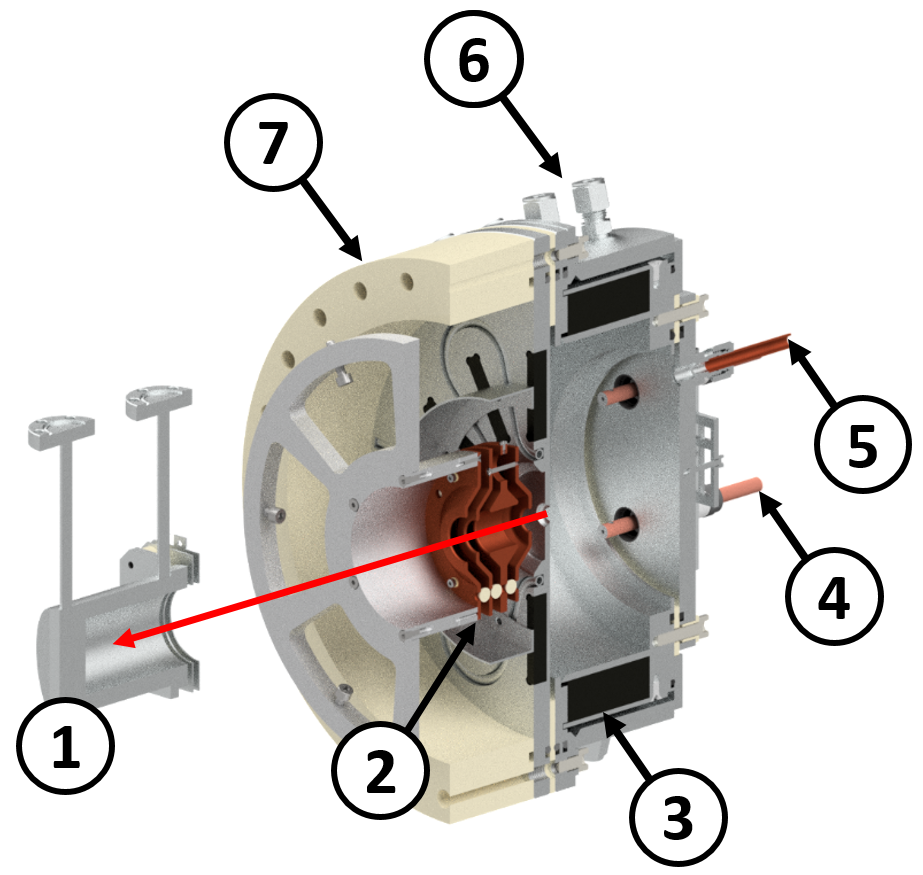}
	    \caption{Cut view of the MIST-1 ion source.
	             1. Faraday cup, 2. Extraction System,
	             3. Permanent magnets (Sm$_2$Co$_{17}$),
	             4. Filament feedthroughs,
	             5. Gas inlet,
	             6. Water cooling fittings,
	             7. Alumina insulator ring.}
	\label{fig:source}
\end{figure}
The baseline design of MIST-1 was described in detail in
\cite{axani:mist1}. In short, it is a filament-driven multicusp ion source
using Sm$_2$Co$_{17}$ permanent magnets for confinement and a 
thoriated (2\%) tungsten filament.
A 3D model of the ion source is shown in Fig. \ref{fig:source} with the 
most important items labeled. The main parameters are listed in the following table.

\begin{table}[!h]
    \centering
    \begin{tabular}{llll}
        \hline
        \textbf{Parameter} & \textbf{Value (nominal)} & \textbf{Parameter} & \textbf{Value (nominal)}\\
		\hline
            Plasma chamber length & 6.5 cm & Plasma chamber diameter & 15 cm \\
            Permanent magnet material & Sm$_2$Co$_{17}$ & Permanent magnet strength 
            & 1.05 T on surface\\
            Front plate magnets & 12 bars (star shape) & Radial magnets & 12 bars \\
            Back plate magnets & 4 bars in 3 parallel rows &
            Front plate cooling & embedded steel tube  \\ 
            Chamber cooling & water jacket & Water flow (both) & (1.5 l/min)\\
            Back plate cooling & none & Filament feedthrough cooling & air cooled heat sink \\
            Filament material & 98\% W, 2\% Th & Filament diameter & $\approx 1.5$ mm  \\
            Discharge voltage & max. 150 V & Discharge current & max. 24 A \\
            Filament heating voltage & max. 8 V & Filament heating current & max. 100 A\\
        \hline
    \end{tabular}
\end{table}

\section{FIRST COMMISSIONING RESULTS}
\begin{figure}
	\centering
		\includegraphics[width=0.4\textwidth]
        {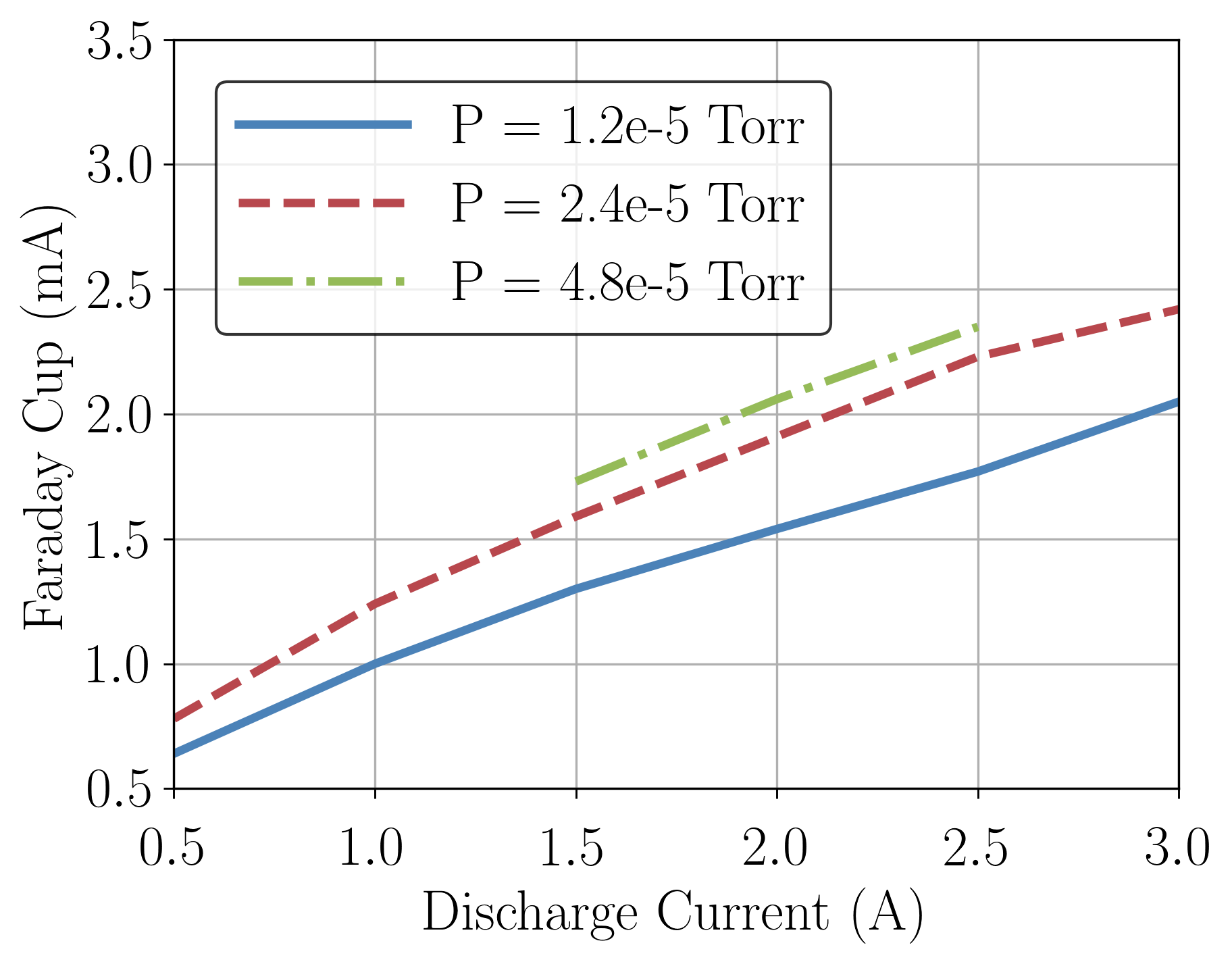}
	    \caption{Total current vs. discharge for different mass flow settings. 
	             The pressure was measured in the chamber shortly after extraction.
	             A clear increasing trend can be seen, suggesting that the upper 
	             limit for total extracted current has not yet been reached.}
	\label{fig:systematic}
\end{figure}

In the first commissioning phase, a thinner (0.4 mm diameter) pure tungsten 
filament was used instead of the nominal filament described in the previous section. Currents were measured in a Faraday cup right after the extraction system (see Fig. \ref{fig:source}), thus not allowing species separation. All reported currents are total extracted currents. 
A first systematic measurement of total extracted current versus discharge current is shown in Fig. \ref{fig:systematic}. A clear increasing trend can be observed that 
suggests the maximum extractable current has not yet been reached.
Indeed, during the accumulated run time of $\approx 30$ hours, the source showed good stability for about 4 hours at a time, reaching a maximum current density of 
16 mA/cm$^2$ (4.6 mA total). This is the maximum that can currently be focused into the
Faraday cup using the einzel lens incorporated in the extraction system, due to power
supply limitations.

\section{SUMMARY AND OUTLOOK}
The MIST-1 multicusp ion source is now operational at the MIT Plasma Science 
and Fusion Center and under commissioning. First results include long-term 
($\ge 4$ hours) stable beams on the order of 4-5 mA total extracted current 
with a 0.4 mm diameter tungsten filament.
First systematic measurements of extracted current as a function of discharge 
voltage and current as well as gas pressure show increasing trends for all three parameters, only limited by source back plate heating and underperforming 
extraction electrode power supplies.
With improved water cooling and new power supplies, we expect to see a significant increase of total extracted current in the next month.
Before the end of the year, the beam line will be extended to include a dipole magnet and two Allison electrostatic emittance scanners (horizontal and vertical) for species analysis and beam quality measurements. Simulations and
design are on the way and a dipole magnet as well as two quadrupole magnets
are on site already.
In the long run, MIST-1 will become part of the RFQ-Direct Injection Project (RFQ-DIP) \cite{winklehner:rfq2}, a test stand for directly injecting 
\htp beams through an RFQ into a compact test cyclotron.

\section{ACKNOWLEDGMENTS}
This work is being supported by NSF Grant No. PHY-1505858 and funding from the Bose Foundation. The authors are very thankful to the MIT Plasma Science and Fusion Center (PSFC) for providing space and utilities to run this experiment and the University of Huddersfield for the lending of equipment.

\bibliographystyle{aipnum-cp}%
\bibliography{Bibliography2}

\end{document}